\documentclass[letterpaper,superscriptaddress,aps,twocolumn]{revtex4-1}
\usepackage{amsmath,amssymb}
\usepackage{bm}
\usepackage{graphicx,color,natbib}
\usepackage{placeins}
\usepackage[dvipsnames]{xcolor}
\usepackage{lipsum}

\newcommand{\beq}{\begin{equation}}
\newcommand{\eeq}{\end{equation}}
\newcommand{\beqa}{\begin{eqnarray}}
\newcommand{\eeqa}{\end{eqnarray}}

\newcommand{\bbeta}{{\bm \bbeta}}

\newcommand{\mum}{\mu\mathrm{m}}
\newcommand{\orient}{\mathbf{\hat{n}}}
\newcommand{\vel}{\mathbf{v}}
\newcommand{\dt}{D_{\mathrm{t}}}
\newcommand{\dr}{D_{\mathrm{r}}}
\newcommand{\EQ}{E_{\mathrm{Q}}}
\newcommand{\EC}{E_{\mathrm{C}}}

\definecolor{strawberry}{rgb}{1.0,0.0,0.5}
\newcommand{\paddyspeaks}[1]{{\color{black} #1}}

\newcommand{\abrahamspeaks}[1]{{\color{black} #1}}

\newcommand{\pecl}{\operatorname{\mathit{P\kern-.08em e}}}

\begin{document}

%\title{Dynamics of Quincke Roller Clusters: Spinning, Orbiting and Flipping}%\paddycomment{ $\leftarrow$ Paddy: happy to change this!}}
\title{Dynamics and Interactions of Quincke Roller Clusters: from Orbits and Flips to Excited States}

\author{Abraham Mauleon-Amieva}
\affiliation{H.H. Wills Physics Laboratory, Tyndall Avenue, Bristol, BS8 1TL, UK}
\affiliation{School of Chemistry, University of Bristol, Cantock's Close, Bristol, BS8 1TS, UK}
\affiliation{Centre for Nanoscience and Quantum Information, Tyndall Avenue, Bristol, BS8 1FD, UK}
\affiliation{Bristol Centre for Functional Nanomaterials, Tyndall Avenue, Bristol, BS8 1FD, UK}

\author{Michael P. Allen}
\affiliation{H.H. Wills Physics Laboratory, Tyndall Avenue, Bristol, BS8 1TL, UK}
\affiliation{Department of Physics, University of Warwick, Coventry, CV4 7AL, UK}

\author{C. Patrick Royall}
\affiliation{H.H. Wills Physics Laboratory, Tyndall Avenue, Bristol, BS8 1TL, UK}
\affiliation{School of Chemistry, University of Bristol, Cantock's Close, Bristol, BS8 1TS, UK}
\affiliation{Centre for Nanoscience and Quantum Information, Tyndall Avenue, Bristol, BS8 1FD, UK}
\affiliation{Gulliver UMR CNRS 7083, ESPCI Paris, Universit\'{e} PSL, 75005 Paris, France}

\begin{abstract}
Active matter systems may be characterised by the conversion of energy into active motion, e.g. the self-propulsion of microorganisms. Artificial active colloids form models which exhibit essential properties of more complex biological systems but are amenable to laboratory experiments. While most experimental models consist of spheres, such as Janus particles, active particles of different shapes are less understood. In particular, interactions between such active colloidal ``molecules'' are largely unexplored. Here, we investigate the motion of active colloidal molecules and the interactions between them. We focus on self-assembled dumbbells and trimers powered by an external electric field. For dumbbells, we observe an activity--dependent behavior of %between
spinning, circular and orbital motion. Moreover, collisions between dumbbells lead to the hierarchical self-assembly of tetramers and hexamers, both of which form rotational excited states. On the other hand, trimers exhibit a novel type of flipping motion that leads to trajectories reminiscent of a honeycomb lattice.
\end{abstract}

\maketitle

%%%%%%%%%% INTRO %%%%%%%%%%

\section{Introduction}

%The hallmark of active matter is the ability to convert energy into motion.
In recent years, much effort has been devoted to investigating the motility of microorganisms, driven, e.g., by flagellae and to the realisation of synthetic active matter by means of diffusophoretic \cite{bocquet2012}, thermophoretic \cite{jiang2010}, field driven \cite{bricard2013} and vibrated particles \cite{deseigne2010}. These active particles exhibit fascinating collective phenomena not found in passive systems, such as flock formation \cite{solon2015,bricard2013}, dynamical clustering and phase separation \cite{buttinoni2013,cates2015,vanDerLinden2019}, anomalous density fluctuations \cite{narayan2007}, vortices \cite{bricard2015} activity-dependent phase behaviour \cite{mauleon2020,sakai2020} and novel dynamical transitions \cite{ravazzano2020}.  While many experiments have focused on self-propelled spherical colloids, e.g. Janus particles, biological microswimmers are often anisotropic, \cite{bechinger2016,elgeti2015,berg2000,lauga2009}.

Many artificial swimmers display a persistent random walk dominated by ballistic runs and rotational diffusion, whereas the locomotion of microorganisms allows adjustments in their trajectories \cite{polin2009,son2013,riedel2005,lauga2006}. Some synthetic particles with motility akin to that of certain biological agents have been obtained, such as an artificial flagellum \cite{dreyfus2005}, rod-shaped \cite{paxton2006}, chiral \cite{ghosh2009}, asymmetric particles \cite{kummel2013} and the balance between run--and--tumble and rotational diffusion can be controlled \cite{karani2019}.

%While interactions between constituent particles drive collective phenomena such as phase behaviour,
In active colloidal systems attention has often focussed on assembly of large numbers of particles,
\paddyspeaks{i.e. the emergence of macroscopic states or active ``phase behavior''} \cite{bricard2013,yan2016,klongvessa2019,mauleon2020}. By contrast, assembly of fixed numbers of \emph{passive} colloids, often through careful control of interactions \cite{glotzer2007} has led to \emph{supracolloidal chemistry} with reaction pathways at the colloidal rather than molecular, level \cite{manoharan2003,hong2006,vanblaaderen2003,chen2011,chen2011a,wang2012,kraft2013,meng2010,zhang2016,snoswell2007,palacci2013,ma2015,denijs2015,niu2017} which %mimic molecular--like
\paddyspeaks{exhibit some aspects of moleculae} interactions \cite{bianchi2006,wang2012}. However, passive colloids exhibit overdamped dynamics, so collisions as such are very different to those that would occur in atomic and molecular systems \cite{brouard,sun2020}. While many active colloidal systems are diffusive at long times, the persistence length of their motion can be many particle diameters, and thus one may enquire as to collisions between active colloids. Activity thus presents a new twist on supra--colloidal chemistry that may be seen as a classical analogue to collisions between atoms and molecules, coupling between them and the formation of excited bound states \cite{brouard,sun2020}.

So far in active colloids and indeed cluster assembly kinetics has been investigated \cite{ginot2018}, interactions between spinning micro--tori \cite{baker2019}, and in simulation predictions have been made for assembly of dimers and trimers of active colloids  \cite{shen2019} and the \paddyspeaks{effect of} torque on assembly of active rods \cite{vandamme2019}. However, in experiments, studies of dynamics of assembled colloidal molecules and their interactions are relatively unexplored.

Here, we present an experimental study on the motion of \emph{active colloidal molecules} \cite{manoharan2003,vanblaaderen2003,lowen2018}. These display a characteristic behaviour different from the directed motility of spheres, in particular circular and jumping motion. To investigate the active colloidal molecules, we exploit the so-called \emph{Quincke} electro-rotation of particles \cite{quincke1896}. Particles are confined between two conductive glass slides $30 \mum$ apart. A dc electric field $E$ is applied to the system, leading to the spontaneous symmetry breaking of the charge distribution at the particle-liquid interface. As a result, rotation at a constant rate emerges from an imposed electric torque acting on the particle. For a rigid sphere near to a substrate, the rotation of the particles is coupled with the translation, giving rise to self-propelled rollers, where the speed $v$ is controlled by the electric field $E$ \cite{bricard2013}. To our knowledge, this is the first time that the self-propulsion of non-spherical rigid particles using the Quincke rotation has been studied.

\begin{figure*}
\centering
\includegraphics[width = 0.99\textwidth]{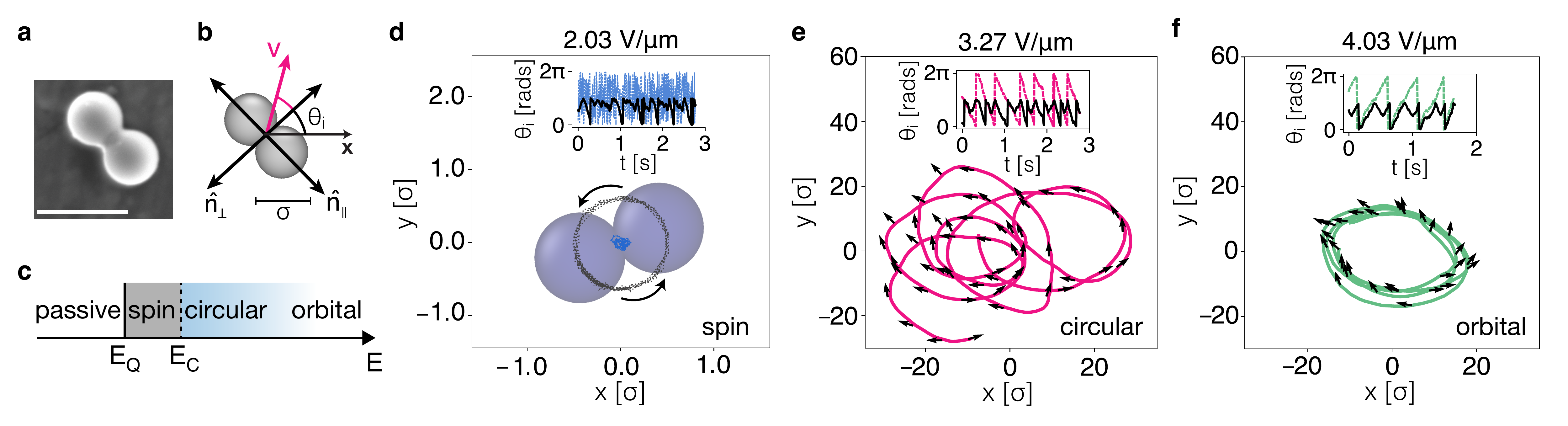}
\caption{\textbf{Spinning, circular and orbital motion of dumbbells.}
\textbf{a.} SEM micrograph of a Quincke dumbbell. Bar represents $5 \mum$.
\textbf{b.} Representation of a dumbbell body-frame. A perpendicular $\bot$ and a longitudinal $\parallel$ orientation $\orient$ with respect to the bond between the two spheres are shown. In addition, the velocity $\vel$ is given by the displacement of the centre-of-mass $\bf{r}$. The angle $\theta_{i}$, corresponding either to the velocity or the orientation, is defined with respect to the reference axis.
\textbf{c.} Field-dependent behaviour of dumbbells. Passive dumbbells become active spinners above $\EQ$, and then circular rollers above $\EC$.
\textbf{d.} A spinning dumbbell at low $E$. Solid line represents the displacement of the centre-of-mass $\bf{r}$, and the dashed line is from the motion of one the sides as the dumbbell spins.
\textbf{e.} Chaotic circular, and \textbf{f.} Orbital motion. Solid lines indicate the displacement of the centre-of-mass, and arrows correspond to the orientation $\orient_{\bot}$. Insets in d,e and f show the time evolution of the angle $\theta_{\vel}$ (dashed lines) and $\theta_{\orient}$ (solid lines).}
\label{figDumbbellsA}
\end{figure*}

We focus on the active motion of colloidal dumbbells and trimers made of two and three spherical particles respectively. A dynamic transition is observed for dumbbells from local spinning to circular and orbital motion as the activity increases. As a result, dumbbells exhibit an increased trajectory radius and a change of their effective translation and rotational motion. In agreement with the description of a Brownian circle swimmer \cite{vanTeeffelen2008}, the self-propulsion direction does not strictly coincide with the dumbbell orientation, resulting in two-dimensional circular trajectories. We observe a dynamical formation of an excited state of spinning tetramers as two dumbbells collide and couple to form a bound state. A more complex formation of hexamers made of three colliding dumbbells is also observed. We find that the spinning motion of tetramers and hexamers is activity-dependent with a coupling between self-propulsion and arrested motion due to steric frustration. Furthermore, for trimers we observe an interesting combination of \emph{in-plane} and \emph{out-of-plane} motion. This corresponds to a jump-diffusion process that evolves the position and orientation of the trimer discontinuously.

%%%%%%%%%%%% RESULTS %%%%%%%%%%%%%%

Our active colloidal molecules are prepared by taking advantage of irreversible attractive interactions between spheres. Briefly, we use polystyrene beads of diameter $\sigma = 3.1 \, \mum$ and polydispersity of 5\%. The initial suspension is %an
aqueous, % medium,
and the \paddyspeaks{colloids are electrostatically} %charged spheres are ionically
stabilised.
To remove ionic stabilizing layers, the particles are washed and transferred to a liquid of low conductivity. This leads to the formation of clusters of different size. Smaller particles are separated from the bigger ones by using centrifugation, and the final suspension is a mixture of single spheres, dumbbells and trimers (see the Supplemental Information for more details \cite{chasingSI}).

%%%%%%%%% DUMBBELLS %%%%%%%%%

\begin{figure*}
\centering
\includegraphics[width = 0.99\textwidth]{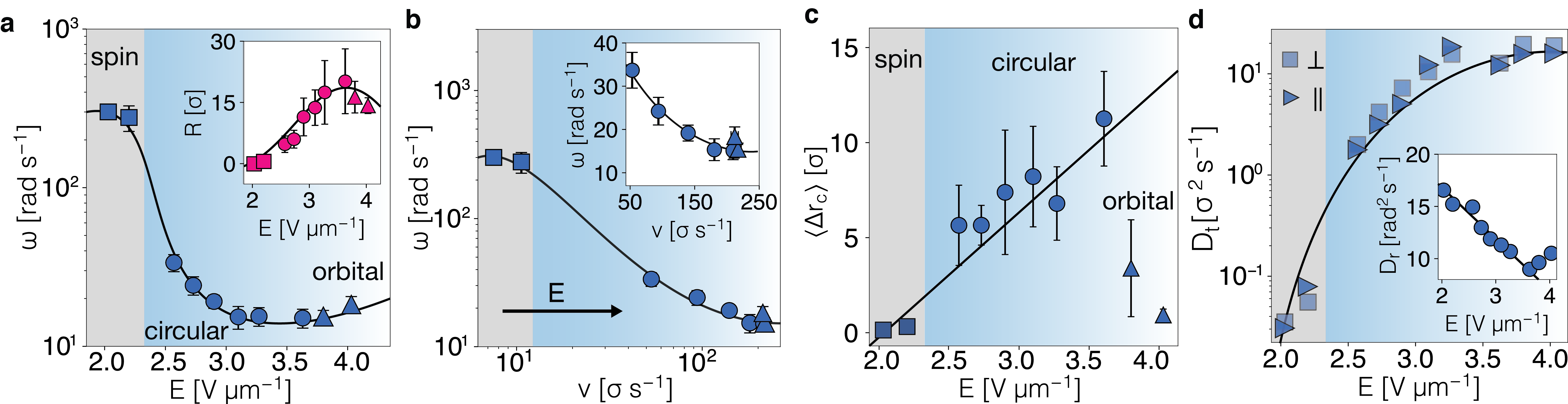}
\caption{\textbf{Dynamics of dumbbells.} \textbf{a.} Angular velocity $\omega$ as function of the field strength $E$. Two \paddyspeaks{regimes} %regions
are identified: spin motion (shaded region) appears with low values of $E$, whereas circular trajectories emerge with the increase of $E$. The same behaviour is observed for the trajectory radius $R$, shown at the inset. \textbf{b.} Angular velocity $\omega$ versus the dumbbell linear speed as the field $E$ increases (arrow). Inset shows the decay of $\omega$ when dumbbells perform circles and orbits. \textbf{c.} Mean displacement of the trajectory central point $\bf{r}_{\rm{c}}$. Symbols in a, b and c are: Spin (squares), circular (circles), and orbital (triangular) motion. \textbf{d.} Effective diffusion coefficients $D$ extracted from the normal and longitudinal mean squared displacements $\langle \Delta (t)^{2} \rangle$ for increasing $E$. The inset displays the values for the effective rotational diffusion coefficients $\dr$ for the same values of $E$.}
\label{figDumbbellsB}
\end{figure*}

\section{Dumbbells: circles and orbits}

We start by describing the active motion of Quincke dumbbells. These are elongated rigid particles with a transverse ($\bot$) and a longitudinal ($\parallel$) orientation $\orient = (\cos \theta, \sin \theta)$, where $\theta$ is the angle formed with respect to a reference axis. Figure \ref{figDumbbellsA}(b) depicts $\orient_{\bot}$ and $\orient_{\parallel}$ with respect to the bond connecting the two spheres. In addition, an angle $\theta_{\vel}$ is given for the displacement. For the motion, we apply a range of field strengths $E \in \{2,4\}\, V\, \mum^{-1}$.
For low values of $E$, i.e. $E < \EQ $ with $\EQ \approx 2 \, V \mum$, we obtain passive dumbbells. Above $\EQ$, a spinning behaviour is observed with constant rate and without a significant displacement of the centre-of-mass $\bf{r}$, as shown in Fig. \ref{figDumbbellsA}(d). The onset of the circular motion occurs at higher values of the applied field, i.e. $\EC \approx 2.5 \, V \, \mum^{-1}$. Finally, at still higher field strengths, the circular motion becomes localised around a central point, giving rise to orbits (see Fig. \ref{figDumbbellsA}(f) and Supplementary Movie 1 in \cite{chasingSI}).

For the circular and orbital trajectories, we find that the dumbbell displacement occurs with a direction $\theta_{\vel}$ close to the transverse orientation $\orient_{\bot}$, different to the directed motion of asymmetric dumbbells controlled by light \cite{yu2018,schmidt2019} and ac fields \cite{ma2015a}. However, such trajectories are a result of the decoupling between the self-propulsion $\bf{v}$ and the dumbbell orientation $\orient_{\bot}$, as indicated by the arrows and insets in Figs. \ref{figDumbbellsA}(d)-\ref{figDumbbellsA}(f). Here, the direction of motion, i.e. clockwise (+) or anti-clockwise(-), is not predefined as in chiral particles \cite{kummel2013}, and thus, the circular motion is given by any slight difference in the sphere size. Following Ref. \cite{vanTeeffelen2008}, the dynamics of non-interacting circle swimmers in two dimensions are given by following the overdamped Langevin equations,
\beq
\begin{aligned}
\dot{\bf{r}} = \beta \bf{D} \cdot [ \it{F} \orient +  \pmb{\zeta}];\\
\dot{\theta} = \beta \dr [\mathbb{T} + \zeta_{\theta} ],
\label{eqCircularLan}
\end{aligned}
\eeq
\noindent
where $\beta = (k_{\rm{B}} \it{T})^{-1}$ is the thermal energy, and $\it{F}\orient$ is an effective internal force representing the self-propulsion. $\bf{D} = \it{D}_{\bot}(\bf{I}-\orient \otimes \orient) + \it{D}_{\parallel}(\orient \otimes \orient)$ is the dumbbell diffusion tensor, where $\it{D}_{\bot}$ and $\it{D}_{\parallel}$ are the transverse and longitudinal translational diffusion coefficients, and $\bf{I}$ is the unit tensor. The rotational dynamics are given by $\dr$, the rotational diffusion coefficient, and $\mathbb{T}$, the effective torque promoting the circular motion on dumbbells. Finally, Gaussian noise terms $\pmb{\zeta}$ and $\zeta_{\theta}$ for the displacement and the orientation are added respectively. Effective diffusion coefficients for both directions are shown as function of $E$ in Fig. \ref{figDumbbellsB}(d). We observe an enhanced translation in both directions during circular and orbital walks. On the other hand, the rotational motion is diffusive, and angular diffusion coefficients are obtained as  \cite{han2006}
\beq
\dr = \langle [\Delta \theta_{\bot}(t)]^{2} \rangle / (2t)
\label{eqDr}
\eeq
for the same values of $E$ (Fig. \ref{figDumbbellsB}(D)).

\textit{Coupling of drive and slipping leads to a non--monotonic angular velocity --- }
Figure \ref{figDumbbellsB}(a) shows the relation of the angular velocity $\omega = | \Delta \theta| / t$ with the applied field $\it{E}$. We observe a non-monotonic decay of $\omega$ as dumbbells perform a circular motion. This behaviour is in good agreement with the response of $\omega$ \abrahamspeaks{against} the increased self-propulsion speed $v$ (Fig. \ref{figDumbbellsB}(b)). Both the increase of $v$ and the decrease of $\omega$ have an impact on the orbit radius $R = v /  |\omega|$. In Fig. \ref{figDumbbellsB}(a) inset we observe a non-monotonic dependence on $v$, with a peak in $R$ at $E \approx 3.6 \, V \mum^{-1}$. For every cycle, $\theta_{v} \in  \{0,2 \pi\}$, we take the displacement of the central point of the trajectory $\Delta \bf{r}_{c}$, which shows the \emph{chaotic} behaviour of the circular motion compared to steady orbits at higher $E$ (Fig. \ref{figDumbbellsB})(c). This is also reflected by the simultaneous increase and decrease of the effective translational and rotational diffusion coefficients in Fig. \ref{figDumbbellsB}(d) respectively. It is worth noting that such behaviour contrasts with the rather steady radius in trajectories of asymmetric active particles \cite{kummel2013}. In such a case, the angular velocity $\omega$ increases linearly with the speed $v$, while $R$ shows a non-dependent behaviour on the self-propulsion. On the other hand, the dependence of $\omega$ and $R$ on $E$ observed here is opposite to the observations of self-propelled spheres in a viscoelastic medium \cite{narinder2018}. Thus, our findings suggest that the emerging circular behaviour of Quincke dumbbells is due to an effective internal torque and to the field-dependent speed $v$.

\begin{figure*}
\centering
\includegraphics[width=0.95\textwidth]{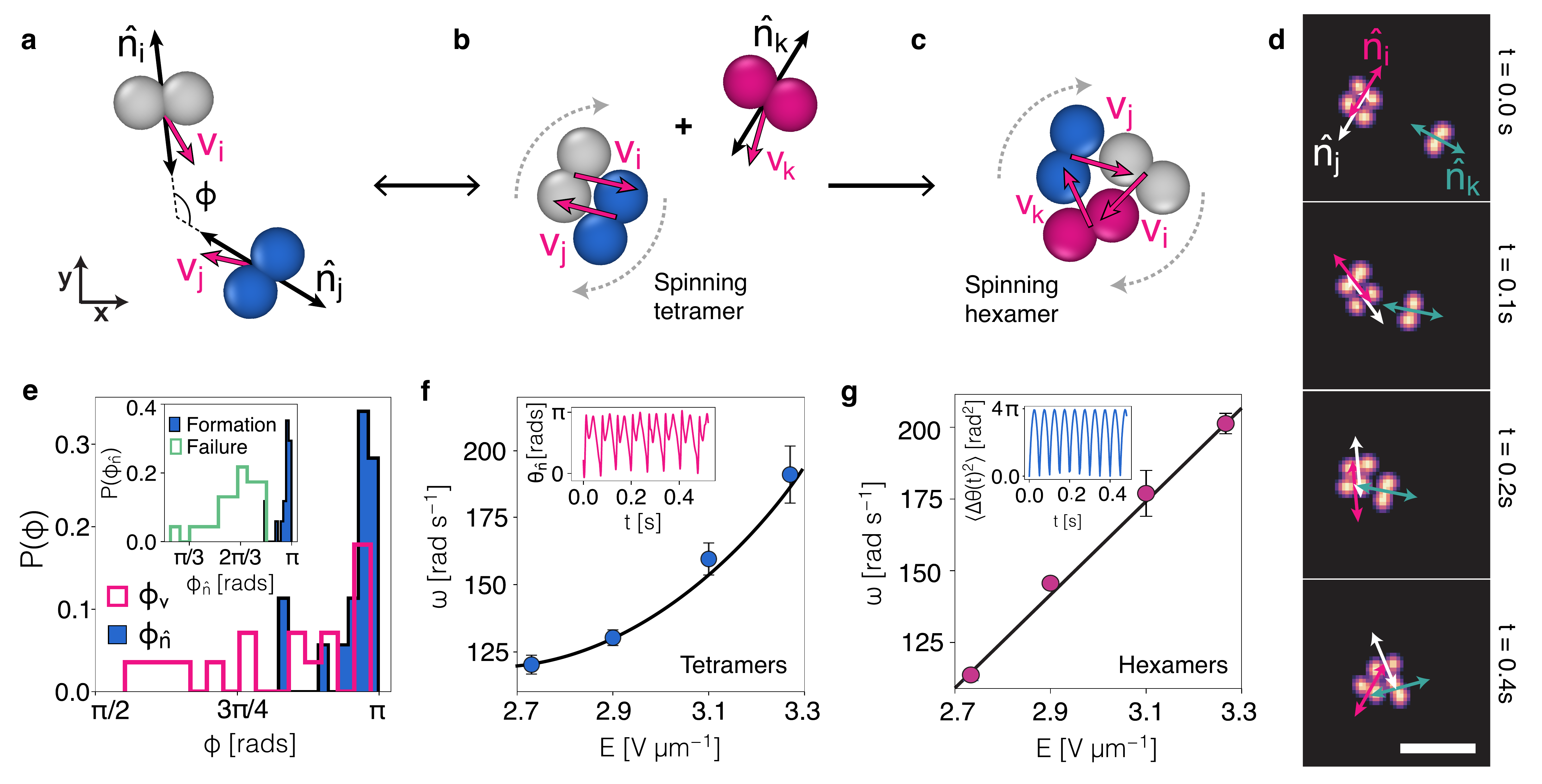}
\caption{\textbf{Dynamical formation of tetramers and hexamers. a.} Active dumbbells performing circular motion are able to collide with a consequent change in their trajectory. We take the angle $\phi_{i}$ made between the orientations $\orient_{ij}$ and velocities $\vel_{ij}$ to characterise the collisions. \textbf{b.} When aligned, two colliding dumbbells form spinning tetramers whose motion results from the dynamical frustration exerted by one dumbbell to the other. \textbf{c.} The dynamical formation of hexamers is possible when a third dumbbell collides with a previously formed tetramer. The resulting spinning motion of hexamers is also attributed to the dynamical frustration of single circular trajectories. \textbf{d.} Formation sequence of a hexamer. A tetramer is previously formed by two dumbbells. A third dumbbell approaches with its orientation $\orient_{k}$ pointing towards the tetramer. Upon collision, the dumbbells rearrange to form a triangular shape as indicated by the orientations $\orient_{ijk}$. \textbf{e.} For dumbbells forming tetramers, the distributions of $\phi_{i}$ indicate that the process is dominated by the dumbbell orientation rather than the velocity. Inset shows the distribution of the orientation angles $\phi$ for successful and unsuccessful formation of tetramers. Bar represents $10 \mum$. \textbf{f-g.} Spinning angular velocities $\omega$ for {f.} tetramers and \textbf{g.} hexamers as function of $E$. Inset in \textbf{f.} is the evolution of the orientation angle $\theta_{\orient}$ as the tetramer spins. Inset in \textbf{g.} shows the mean angular displacement $\langle \Delta \theta(t)^{2} \rangle$ of a spinning hexamer.}
\label{figTetraHex}
\end{figure*}

\begin{figure*}
\centering
\includegraphics[width=0.99\textwidth]{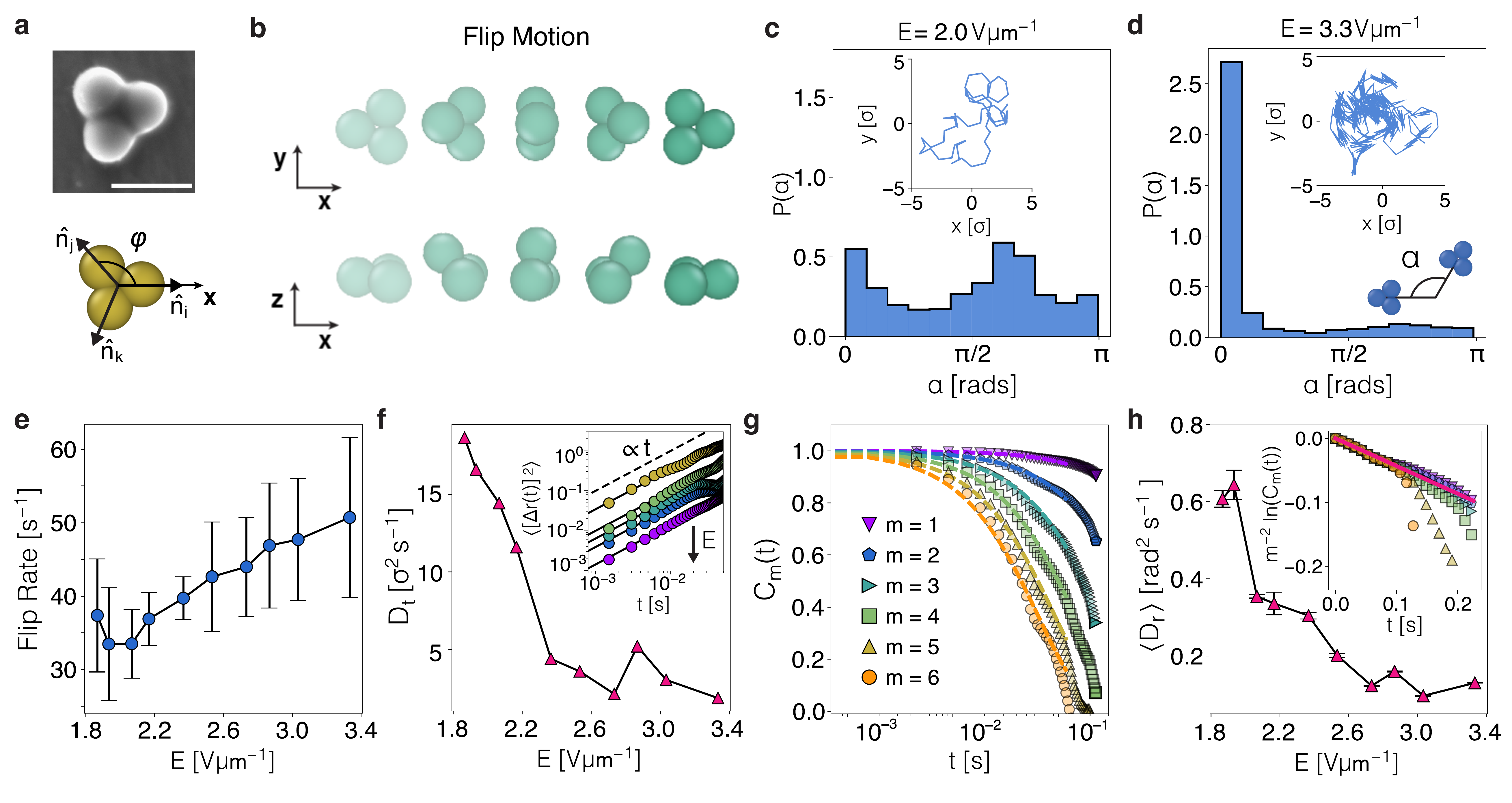}
\caption{\textbf{Quincke trimers. a.} \textit{Top:} SEM micrograph of a Quincke trimer. Scale bar is $5 \mum$. \textit{Bottom:} Trimer body-frame. The orientation $\orient$ of each vertex is given by an angle $\phi$ formed with respect of a reference axis and each vertex position.
\textbf{b.} Schematic representation of the flip motion performed by active trimers. Every jump corresponds to a leapfrogging mechanism of one vertex over the opposite side of the trimer.
\textbf{c.} Distribution of flip angles $\alpha$ for a trajectory $E \approx 2 \, V \mum^{-1}$, shown at the inset.
\textbf{d.} A trimer trajectory dominated by flips at $E \approx 3.33 \, V \mum^{-1}$ (inset) shows a strong distribution of $\alpha \rightarrow 0$.
\textbf{e.} Flip rate as function of the electric field strength $E$.
\textbf{f.} Effective translational diffusion coefficients $\dt$ obtained from filtered trajectories. Inset shows diffusive mean squared displacements. Arrow indicates increase of $E$. \textbf{g.} Reorientational time correlation functions for six different ranks $m$ as defined in the main text. Symbols are obtained from experimental trajectories at $E \approx 2 \, V \mum^{-1}$, and dashed lines are fittings from Eqn~\eqref{eqRcorrelation}. Inset in \textbf{h} displays the collapsed scaled functions for the same data shown in \textbf{g}. Solid line is a fitting using the mean rotational diffusion coefficient $\dr$ extracted from the fits in \textbf{g}. \textbf{h.} Effective rotational diffusion coefficients $\dr$ versus the applied electric strengths.}
\label{figTrimers}
\end{figure*}

%%%%%%% TETRAMERS / HEXAMERS %%%%%%%%%%%

\section{Dumbbell collisions: hierarchy of ``excited states''}

Having a dilute suspension of dumbbells performing a circular motion, e.g. at $E \in \{2.5,3.5\} \,V \mum^{-1}$, we observe collisions between dumbbells that lead to the formation of tetramers and more complex hexamers. Figure \ref{figTetraHex}(a) shows the sequential formation of these ``excited states''. First, isolated dumbbells, with orbits of sufficient length $R$ (see inset in Fig. \ref{figDumbbellsB}(a)), collide and interrupt each other's motion. If the collision is successful in terms of \emph{alignment}, the result is a spinning tetramer of rhomboidal shape (see Fig. \ref{figTetraHex}(b) and Supplementary Movie 2 in Ref. \cite{chasingSI}). We argue that this ``excited state'' is due to the dumbbells colliding and being unable to move past one another following the collision. In other words, the particle geometry enables dynamical self--trapping similar to motility--induced phase separation \cite{cates2015} which here results in a bound state.

For colliding dumbbells, we measure the angle $\phi$ made between the  orientations $\orient^{\bot}_{ij}$ and velocities $\bf{v}_{\it{ij}}$ prior to collision and tetramer formation. Figure \ref{figTetraHex}(e) shows the distributions of the $\phi_{\orient,\vel}$ angles. While the displacements exhibit a broader distribution, the successful formation of tetramers is governed by the dumbbell orientation. That is to say, this formation of tetramers is achieved by an effective alignment, i.e. the orientation angle $\phi \rightarrow \pi$ and the displacements $\mathbf{v}_{i} + \mathbf{v}_{j}=0$. In addition, we compare successful formation against other dumbbell collision, confirming the strong dependence on orientation (see inset in Fig. \ref{figTetraHex}(e)). If unperturbed, tetramers spin at a constant angular speed $\omega$ and without significant displacement of the centre-of-mass $\Delta \bf{r}$. \paddyspeaks{The rotation results from the torque as the centre of propulsion from each dumbbell is not aligned with the centre of mass of the tetramer (Fig. \ref{figTetraHex}(b)).} Otherwise, any significant change in the orientation $\orient_{\bot}$ promotes tetramer breaking and reversal to the circular motion of dumbbells. Figure \ref{figTetraHex}(f) depicts the spinning speed $\omega$ of tetramers as function of $E$. We observe increasing $\omega$ with the field as a result of the enhanced self-propulsion $v$ (Fig. \ref{figDumbbellsB}(b)).

\textit{Hexamers: unstable excited states --- }
In a more complex scenario, spinning hexamers form due to the self-trapping of three dumbbells. For this, an additional dumbbell collides with a pre-existing tetramer. A triangular hexamer results from the local rearrangement of dumbbells, as represented in Figs. \ref{figTetraHex}(b) and (c) (see also Supplementary Movie 4). Similar to tetramers, it is likely that the process is governed by the dumbbell orientation $\orient_{\bot}$. In Fig. \ref{figTetraHex}(d) we show an experimental formation sequence of a hexamer, where the orientations $\orient_{i,j,k}$ for each dumbbell are highlighted. We find few spinning hexamers using the same values of $E$ as for tetramers. The spinning speed shows a linear increase with $E$, suggesting a stronger coupling of the individual self-propulsion speeds $v$. In contrast to tetramers, the breaking of hexamers shows no reversion as any deviation of the individual orientation $\orient$ leads to the segregation of dumbbells (Supplementary Movie 3). The break--up of the hexamer in Supplementary Movie 3 underlines the complex hydrodynamic couplings in colloidal system under DC fields \cite{yeh2000}.

We emphasize that the collision processes here are different to the phoretic and hydrodynamic interactions of Janus particles \cite{bocquet2012} and Quincke rollers \cite{mauleon2020}. The formation of tetramers and hexamers is due to dynamical self-trapping, akin to systems displaying phase separation \cite{buttinoni2013}. \abrahamspeaks{Therefore, the active pathway of hierarchical states is distinctive from the ones observed in systems with induced interactions \cite{zhang2016, yan2012}.}

%%%%%%% TRIMERS %%%%%%%%%%

\section{Trimers: flipping on a honeycomb lattice}

We now proceed to describe the active motion of trimers, which are rigid assemblies of three particles (Fig. \ref{figTrimers}(a)). Similar to dumbbells, these triangular particles are prone to an effective reorientation due to activity. However, the hallmark of trimers is their distinctive \emph{flip} motion. This consists of a jump performed by one vertex over the opposite side, as represented in Fig. \ref{figTrimers}(b) (see Supplementary Movie 5 in \cite{chasingSI}). This corresponds to a novel class of \emph{Quincke flippers}. Such a jump effectively instantaneously rotates the orientation by $\pi$ about an axis parallel to the triangle side, and displaces the centre-of-mass $\mathbf{r}$ by a distance $\ell$ perpendicular to this axis. Assuming non-slip conditions, $\ell \approx 0.7 \sigma$. In between flips, there is continuous (possibly diffusive) evolution of position and orientation. In the absence of evidence to the contrary, it seems reasonable to assume that these two types of motion occur independently.

In principle, the orientation of the trimer is specified by three Euler angles $(\varphi,\theta,\psi)$. The first angle, $\varphi$, is between the body-fixed and space-fixed $\mathbf{x}$ axes; we take the body-fixed $\mathbf{x}$ axis to point from the trimer center towards a vertex, see Fig.~\ref{figTrimers}(a). The second angle, $\theta$, is a rotation about the body-fixed $\mathbf{x}$ axis. This takes values $0$ and $\pi$ in the unflipped and flipped state, respectively, and, within the instantaneous flip approximation, these are the only values of interest. The third angle $\psi$ may be taken to be zero. Assuming that the dynamics do not depend on the flip state, we may focus on the angle $\varphi$ alone.

For a trimer with particle centres in an equilateral geometry, a flip may be represented as $\varphi\rightarrow\varphi+\Delta\varphi$, $\mathbf{r}\rightarrow\mathbf{r}+\Delta\mathbf{r}$,
where $\Delta \varphi =  \pm\pi/3, \pi$ corresponding to the three possible flip directions, and
\beq\label{eqn:flipgeometry}
\Delta \mathbf{r} = \ell [\cos ( \varphi - \Delta\varphi ),\sin (\varphi - \Delta\varphi )] .
\eeq
In the absence of motion between flips, the centre $\mathbf{r}$ of each trimer would explore the vertices of a two-dimensional honeycomb lattice. Successive flips may or may not be correlated, regarding the time intervals between flips and/or the choice of successive flip directions. The simplest model for the motion between flips is that the trimers translate and rotate diffusively, obeying
\beq
\dot{\mathbf{r}} = (\dot{x},\dot{y}) = \sqrt{2 \dt}(\zeta_{x},\zeta_{y})
\quad\text{and}\quad \dot{\varphi} = \sqrt{2 \dr}\zeta_{\varphi},
\eeq
where $\zeta_{x}$, $\zeta_{y}$, and $\zeta_{\varphi}$ are independent delta-correlated stationary Gaussian processes with zero mean, and $\dt$ and $\dr$ are the translational and rotational diffusion coefficients. An active (velocity) contribution might be added to these equations, but such a term would imply some breaking of triangular symmetry.

Figures \ref{figTrimers}(c) and \ref{figTrimers}(d) display experimental trajectories of a trimer performing a random walk at different activities. In contrast to active spheres and dumbbells, flippers show reduced displacement $\Delta \bf{r}$ due to the symmetry in jumps. For the field range applied here, i.e.\ $E \in \{1.8,3.4 \}\,V \mum^{-1}$, we find that the motion of trimers is dominated by flips. To characterise any correlation of the flips we define an angle $\alpha$ made by two successive displacements of the centre-of-mass (see the diagram in Fig. \ref{figTrimers}(d)). Having uncorrelated flips without rotation in between, the angle made by $\Delta \bf{r}$ takes possible values $+\pi/3$, $-\pi/3$, or $\pi$ occurring with equal probability. For $\alpha$ (as defined in Fig. \ref{figTrimers}(d)), the angles for the three previous cases are $2\pi/3$, $2\pi/3$ (again), and 0. The bimodal distribution in Fig. \ref{figTrimers}(c) reflects this behavior, with a larger peak at $\alpha \approx 2\pi/3$. Upon increasing $E$, we observe weakening of the bimodal nature of the distribution as the distribution of $\alpha$ shifts to $0$, showing an enhanced anisotropic motion, as displayed in Fig. \ref{figTrimers}(d). Also, the increase of $E$ yields an increasing flip rate, as shown in Fig. \ref{figTrimers}(e). The increasing correlation of flips might be given by any small asymmetry in the shape, i.e.\ spheres of different size, that together with the increased field result in linear regions of the trajectory (see inset in Fig. \ref{figTrimers}(d)).

With the above observations in mind, it is possible to devise a dynamical model of the trimer, based on continuous (possibly diffusive) motion, punctuated by instantaneous flips (possibly incorporating the correlations just discussed), to compare with experiments. This model is discussed further in the supplementary information \cite{chasingSI}. Here, though, we focus on isolating the continuous motion, and determining whether it is diffusive. Using the facts that the flips are rapid, and produce displacements $\Delta\mathbf{r}$ and $\Delta\varphi$ which approximately satisfy Eq.~\eqref{eqn:flipgeometry}, it is possible to remove the effects of the flips from the experimentally observed trajectories, leaving just the continuous evolution of $\mathbf{r}(t)$ and $\varphi(t)$. We refer to these as \emph{filtered} trajectories. We emphasize that this is an artificial procedure, only likely to be successful if the two types of motion are independent.

Mean-squared displacements of filtered trajectories obtained at different field strengths $E$ are shown in the inset of Fig.~\ref{figTrimers}(f). The curves can be fitted by $\langle \Delta r(t)^2\rangle = 4\dt t$, and values for the effective diffusion coefficients $\dt$ are shown as a function of $E$ in Fig.~\ref{figTrimers}(f).

Reorientation in the plane is analysed by means of time correlation functions of $\varphi$,
again extracted from the filtered trajectories
\beq
    C_{m}(t) = \langle \cos m \Delta \varphi (t) \rangle = \exp(-m^{2}\dr t)
    \label{eqRcorrelation}
\eeq
where $\Delta \varphi (t) = \varphi(t) - \varphi(0)$, the change in angle due to non-flip motion only, $m$ is the rank, and the last expression is expected for pure rotational diffusion with coefficient $\dr$ \cite{lynden-bell1980,*lynden-bell1985}. In Fig.~\ref{figTrimers}(g) we show specimen results for $m\leq 6$ at one value of $E$. In the inset of Fig.~\ref{figTrimers}(h), the same data collapses onto a single curve $m^{-2} \ln C_{m}(t)$ vs $t$, allowing an estimate of $\dr$. Rotational diffusion coefficients $\dr$ as a function of $E$ are shown in Fig.~\ref{figTrimers}(h). Translational and rotational diffusion seems to satisfactorily describe the motion between flips. We find decay of both $\dt$ and $\dr$ as we increase $E$. It is important to recognize that perfect separation of flips and diffusion may be impractical (in reality, there is a distribution of flip distances and directions, and the trimers are not perfectly equilateral triangles). Hence, the measured ``diffusion'' coefficients may effectively include residual contributions from the flips. If this is the case, then the decrease in $\dt$ and $\dr$ with increasing $E$ might be connected with the increased importance of the $\alpha=0$ peak, i.e.\ correlated forward and backward flips, at higher $E$,
indicated in Fig.~\ref{figTrimers}(d).

\section{Discussion and Conclusions}

In summary, we have investigated active motion of dumbbells and trimers powered by Quincke rotation. The circular motion and the jumping behaviour of these non-spherical particles is markedly different from that
observed in rolling colloids, leading to a new class of Quincke dumbbells and flippers. For both cases, the behaviour is controlled by the applied electric field. The motion of dumbbells observed experimentally at intermediate activity is in agreement with the theoretical description of a circle swimmer \cite{vanTeeffelen2008}.

Such motion transforms into the emergence of spinning tetramers and hexamers as dumbbells collide with each other. This corresponds to an active \emph{supracolloidal} formation of excited bound states arising from the dynamical aggregation of dumbbells. The persistence length inherent in the motion of Quincke Rollers leads to a dependence on the trajectories of the incoming dumbbells, and in particular on the collision angle $\phi$ (Fig. \ref{figTetraHex}). Such a trajectory dependence is absent from the overdamped dynamics of passive colloidal systems \cite{ivlev,hong2006,chen2011}, and indeed the exotic excited states we find are reminiscent of long--lived complexes formed by collisions between molecules but at the colloidal lengthscale and of course with classical interactions \cite{brouard,sun2020}.

Moreover, we have shown the flip behaviour of trimers, described by means of a jump-diffusion model. Although details for the self-propulsion active forces are absent, the approach seems promising. Therefore, our findings provide novel active models to perform experiments with colloids. These might be beneficial for the investigation of different types of motion as encountered in nature, as well as for the design of non-equilibrium self-assembly routes and to provide a readily observable classical analogue of collisions and excited states in molecules.

We thank Mike Ashfold, Olivier Dauchot, Jens Eggers, Tannie Liverpool, Andrew Orr-Ewing, Anton Souslov, Zorana Zoravcic for helpful discussions. C.P.R. would like to acknowledge the European Research Council under the FP7/ERC Grant Agreement No. 617266 NANOPRS. A.M.-A. thanks CONACyT and EPSRC EP/T031077/1 for support.

% \abrahamspeaks{M. A. ....}
% \mikespeaks{I don't think that I have anything to add to this section.}

\bibliographystyle{naturemag}
\bibliography{chasing19ArXiV,chasingCantFaceFightingWithZotero}

\end{document}